\documentclass[preprint,aps,prb,showpacs]{revtex4}
\usepackage[dvips]{graphicx,epsfig}
\usepackage{amsmath,amssymb}

\begin{document}

\title{Observation of Andreev reflection in the $c$-axis transport
of Bi$_2$Sr$_2$CaCu$_2$O$_{8+x}$ single crystals near $T_c$ and
search for the preformed-pair state}

\author{Hyun-Sik Chang}
\author{Hu-Jong Lee}
\thanks{Corresponding author: Hu-Jong Lee; Department of Physics,
Pohang University of Science and Technology, Pohang 790-784,
Korea; telephone/fax (+82-54-279-2072/+82-54-279-3099); e-mail
(hjlee@postech.ac.kr)}
\affiliation{Department of Physics, Pohang
University of Science and Technology, Pohang 790-784, Republic of
Korea}
\author{Migaku Oda}
\affiliation{Department of Physics, Hokkaido University, Sappore
060-0810, Japan}

\begin{abstract}
We observed an enhancement of the $c$-axis differential
conductance around the zero-bias in
Au$/$Bi$_2$Sr$_2$CaCu$_2$O$_{8+x}$ (Bi2212) junctions near the
superconducting transition temperature $T_c$. We attribute the
conductance enhancement to the Andreev reflection between the
surface Cu-O bilayer with suppressed superconductivity and the
neighboring superconducting inner bilayer. The continuous
evolution from depression to an enhancement of the zero-bias
differential conductance, as the temperature approaches $T_c$ from
below, points to weakening of the barrier strength of the
non-superconducting layer between adjacent Cu-O bilayers. We
observed that the conductance enhancement persisted up to a few
degrees above $T_c$ in junctions prepared on slightly overdoped
Bi2212 crystals. However, no conductance enhancement was observed
above $T_c$ in underdoped crystals, although recently proposed
theoretical consideration suggests an even wider temperature range
of enhanced zero-bias conductance. This seems to provide negative
perspective to the existence of the phase-incoherent preformed
pairs in the pseudogap state.
\end{abstract}

\pacs{73.40.-c, 74.50.+r, 74.72.Hs, 74.80.Fp}


\maketitle

\section{Introduction}

It is now widely accepted that the dominant order parameter (OP)
of high-$T_c$ superconductors (HTSC's) has a $d_{x^2-y^2}$-wave
($d$-wave) symmetry,\cite{DJ,twin,Ng,Wei} along with other minor
components.\cite{twin,Covington,Mossle} The existence of nodes and
the azimuthal phase alternation of the $d$-wave OP in the
superconducting Cu-O bilayers leads to novel phenomena in the
electrical transport such as the formation of $\pi$-junctions
\cite{DJ,twin,Schulz} and zero-bias conductance
peak.\cite{Ng,Wei,Tanaka} In general, the zero-bias conductance
peak is observed in normal-metal$/$HTSC junctions with the
tunneling direction along the plane of a Cu-O bilayer. It results
from a coherent interference of incident and reflected
quasiparticles with energies lower than the superconducting gap of
a HTSC electrode near the junction interface. The coherent state
in the $d$-wave superconductor, which is called the Andreev bound
state (ABS), occurs when the quasiparticles in the Cu-O bilayer
experience sign changes of OP upon reflection at the interface.
Thus, the ABS effect is more pronounced for a junction which is
normal to one of the node directions of the $d$-wave OP. A
resonant tunneling of quasiparticles in the normal-metal electrode
to the ABS of the HTSC crystal leads to the zero-bias conductance
peak. In contrast to the conventional Andreev reflection (AR)
effect,\cite{Andreev,BTK} where maximum zero-bias conductance
enhancement is obtained for negligible interfacial barrier
strength, a finite scattering barrier at the interface is
essential for the observation of the zero-bias conductance peak
through the ABS.\cite{Tanaka}

The $c$-axis electrical transport properties of the HTSC's can be
well understood in terms of $c$-axis stacking of intrinsic
Josephson junctions (IJJ's), where the Cu-O bilayers serve as thin
superconducting electrodes and the layers in-between as insulating
layers. The $c$-axis tunneling current-voltage characteristics
(IVC) of the HTSC's such as Bi$_2$Sr$_2$CaCu$_2$O$_{8+x}$ (Bi2212)
exhibit the features of stacked DID
($d$-wave-HTSC/insulator/$d$-wave-HTSC) IJJ's at temperatures well
below the superconducting transition temperature $T_c$.\cite{IJJ}
However, the insulating layer acts as a strong interfacial
scattering barrier and thus hinders the observation of a
conventional AR along the $c$-axis direction in HTSC junctions.
Moreover, the quasiparticles which are reflected from the
transverse boundary of Cu-O bilayers experience no sign change and
thus the observation of the zero-bias conductance peak by the
formation of the ABS in the $c$-axis tunneling cannot be
expected.\cite{Tanaka} In this study, we report observation of the
conventional AR in Bi2212 single crystals {\it for the $c$-axis
transport} at temperatures very close to the bulk superconducting
transition temperature $T_c$. In our previous work,\cite{Kim} it
has been shown that the superconductivity of the surface Cu-O
bilayer in a Bi2212 single crystal is suppressed when the surface
layer is in contact with a normal-metal electrode. Such
suppression leads to the formation of a natural NID junction at
the crystal surface, the ``surface junction'', in the temperature
range $T'_c<T<T_c$, where $T'_c$ and $T_c$ are the transition
temperatures of the surface and inner Cu-O bilayers, respectively.
In this surface NID junction, N represents the surface Cu-O
bilayer with suppressed superconductivity and the normal-metallic
Au electrode, I is the non-superconducting layers between adjacent
Cu-O bilayers, and D is the neighboring inner superconducting Cu-O
bilayers with $d$-wave OP symmetry. As the temperature is raised
and approaches $T_c$, the differential conductance near the
zero-bias voltage of the surface junction develops from
tunneling-like depression to an AR-like enhancement in a
continuous manner. Such development points to gradual weakening of
the barrier strength between adjacent Cu-O bilayers with
increasing temperature. We show that the transitional feature can
be confirmed numerically, at least in a qualitative level, using
the formalism by Blonder, Tinkham, and Klapwidjk (BTK) for
$d$-wave symmetry.\cite{Tanaka}

Recently, it has been proposed by Choi, Bang, and Campbell
\cite{Choi} that, if the pseudogap state above $T_c$ in the HTSC's
is described in terms of the phase-incoherent preformed
pairs,\cite{Emery} the AR can take place from such preformed
pairs. They suggest that experimental observation of the resulting
conductance enhancement in an $ab$-planar normal-metal/HTSC
junction would provide unambiguous confirmation of the preformed
pairs in the pseudogap state. Along with the suggestion of Ref. 13, 
we utilized the appearance of the AR effect {\it along the c-axis
direction} near $T_c$ to investigate the existence of the
preformed pairs in the pseudogap state. We observed the zero-bias
conductance enhancement (ZBCE) in junctions prepared on the
surface of as-grown overdoped Bi2212 single crystals over
a-couple-of-degrees temperature window even above $T_c$. However,
no appreciable ZBCE has been observed above $T_c$ in underdoped
crystals, although theoretical
consideration of Ref. 13 
suggests an even wider
temperature window of ZBCE. Our result is consistent with the
recent observation of the nonexistence of the AR effect {\it in
ab-planar} normal-metal/HTSC junctions by Dagan {\it et al.}
\cite{Dagan} and seems to provide negative perspective to the
existence of the preformed pairs in the pseudogap state.

\section{Experiment}

In this study, we used Bi2212 single crystals of different doping
levels. As-grown overdoped crystals were prepared by the
conventional solid-state-reaction method. Crystals in the
underdoped level were grown first by traveling solvent floating
zone methods. Reducing doping level was done by annealing the
crystals in a low-concentration ($\sim$0.1$\%$) oxygen gas mixed
with nitrogen gas. The superconducting transition temperatures of
the as-grown overdoped and underdoped crystals used were about 90
K and 83 K, respectively. For mesa fabrication,
a-few-thousand-\AA-thick Au film was deposited first on the
surface of a freshly cleaved single crystal. Mesa structure was
then formed by photolithographic micropatterning and ion-beam
etching. Additional process of photolithographic patterning, metal
deposition, and lift-off was done to attach contact leads to the
small mesa area. Details of the mesa fabrication are described
elsewhere.\cite{Kim,Kim2} Finally, the central mesa (refer to the
insets of Fig. 1) of typical lateral dimension of 10$\times$45
$\mu$m$^2$ was divided into equal halves by further etching it
with the contact leads as masks. The dimension of each smaller
mesa thus prepared in the central mesa was 10$\times$13
$\mu$m$^2$. We adopted the configurations in the left and right
insets of Fig. 1 for the four- and three-terminal measurements,
which were used to probe the lower and upper stacks of intrinsic
junctions in the central mesa, respectively. The four-terminal
configuration was used to monitor the intrinsic properties of the
junctions, while the three-terminal configuration prepared on the
same crystal was used to monitor the properties including the
surface junction. The total thickness of the central mesa was
usually less than 200 {\AA} and the upper stack contained about
5-6 intrinsic junctions. The temperature dependence of the
$c$-axis resistance $R_c(T)$ and the differential conductance
$dI/dV$ were obtained by standard ac lock-in technique.

\section{Results and Discussion}

Fig. 1 shows the typical $c$-axis tunneling resistance $R_c(T)$ of
a mesa (UD2) fabricated on the surface of an underdoped Bi2212
single crystal in a three-terminal configuration. The bulk
transition temperature $T_c$ should be close to the temperature of
maximum $R_c(T)$ curve, that is 82.5 K. As Kim {\it et al.}
reported earlier \cite{Kim} the superconductivity of the surface
Cu-O bilayer of a Bi2212 single crystal, which is in contact with
normal-metallic Au electrode, is significantly suppressed. The
resulting formation of the surface NID junction is apparent as
illustrated in Fig. 1 by the strongly semiconducting temperature
dependence of the $R_c$ curve below $T_c$. For the mesa UD2 in the
underdoped regime the superconducting transition of the surface
Cu-O bilayer was not observed down to $\sim$4.2 K, which was in
contrast to previous observations in slightly overdoped
crystals,\cite{Kim,Kim2,Doh1} where the surface bilayers usually
showed the superconducting transition at $T_c' \approx30-40$ K. We
believe that, in the case of underdoped crystals, the reduced hole
doping of the surface Cu-O bilayer \cite{pTc,Mossle2} contributes
to the suppression of the superconductivity in addition to the
proximity-induced suppression.

A set of differential conductances ($dI/dV$) of the mesa UD2 for
various temperatures are shown in Fig. 2. The conductance was
measured {\it in the first branch} of the IVC, which was from the
surface junction of the mesa. The $dI/dV$ curves at the lower
temperature range in Fig. 2(a) display distinct gap-like features.
An intrinsic Josephson junction in the supercurrent branch of its
IVC usually switches prematurely to the quasiparticle branch
before the bias voltage reaches the gap edge. The sudden jumps of
the differential conductance at finite voltages in Fig. 2(a) were
caused by this premature switching, where the Josephson coupling
of the inner junctions broke down before the gap edge of the inner
Cu-O bilayer in the surface NID junction was reached. More details
of the situation are illustrated in the inset of Fig. 2(b) for
$T\approx$ 65.2 K, where the IVC for biases below 16.5 mV and
above 27 mV represent the first and the second quasiparticle
branches, respectively. The shaded area in-between approximately
marks the region of the voltage jump in the IVC. As the
temperature is raised the depression of the differential
conductance near the zero bias becomes shallower (Fig. 2(a)), and
the conductance gradually exhibits an zero-bias enhancement (Fig.
2(b)) between 75.9 K and 79.6 K for this mesa. However, as shown
in Fig. 2(b), in the middle of the superconducting transition
slightly above 79.6 K, the magnitude of the zero-bias peak
decreases with increasing temperature and finally vanishes near
$T_c$. The oscillations of the $dI/dV$ curve with rather regular
periodicity at $T\approx$75.9 K are believed to be caused either
by the fluctuating Josephson coupling in the inner junctions
underneath the surface junction near $T_c$ or by the coherent
interference of boundary-reflected quasiparticles in the
normal-metal electrode of the surface junction.\cite{Rowell,
Nesher}

The appearance of the zero-bias enhancement of the differential
conductance of this {\it c-axis tunneling} at temperatures just
below $T_c$ is very surprising. It is in clear contrast to most of
earlier results of $c$-axis tunneling measurements in
HTSC's,\cite{Wei,cat1,cat2,Miya1,Miya2} where at all temperatures
explored, even very close to $T_c$, only gap-like features have
been observed. A very few cases of conductance enhancement in the
$c$-axis transport were reported \cite{Ng,Ishibashi} but with much
less pronounced peaks than those observed in this study, so that
earlier observation of the c-axis ZBCE was accounted for by
leakage {\it ab-plane tunneling} through some surface defects.

The transition from gap-like depression to enhancement of the
differential conductance was observed earlier in STM measurements
on conventional superconductors.\cite{Sri,Chuang} By reducing the
spacing between the normal STM tip and the superconductor surface
the conduction characteristics of the system turned from
tunneling-like to a weak-link behavior until the direct contact
was made between the STM tip and the superconductor surface. The
barrier strength parameter $Z$ decreased correspondingly, along
with the decrease of the tunneling resistance. Also, in a junction
of Au$/$Bi$_2$Sr$_2$CuO$_6$(Bi2201)$/$Bi2212 single crystal, which
was fabricated by sequential deposition of Bi2201 and Au films on
a Bi2212 single crystal, Matsumoto {\it et al.} \cite{SAR}
observed excess zero-bias conductance with the temperature
approaching $T_c$ from below. It was argued that the observed
excess zero-bias conductance was caused by the conventional AR
effect, where the effective $Z$ value was reduced with increasing
temperature by the semiconducting temperature dependence of the
Bi2201 film. We suppose that the conventional AR effect with
decreasing effective $Z$ value slightly below $T_c$ was also
responsible for the observed ZBCE in our measurements. One should
note, however, that the increase of the resistance below $T_c$ in
Fig. 1 was caused mainly by the temperature dependence of
$-{\partial f(E-eV)}/{\partial (eV)}$ in Eq. (3) in the zero-bias
limit, $V=0$. Thus, the additional weak temperature dependence of
$Z$ should be determined from a careful fitting as described
below. The possibility of an AR between Au layer and surface Cu-O
bilayer can be ruled out, since the latter is non-superconducting
at all temperatures under consideration. We assume that the reason
for the scarcity of observation of the ZBCE in the $c$-axis
transport
measurements of HTSC's except for the cases of Refs. 3 and 26 
was mainly because most of the
previous measurements were focused on the temperature range
sufficiently lower than $T_c$ with resultant strong barrier
potential.

In order to confirm the possible appearance of the AR effect in
the $c$-directional conduction we followed the extended BTK
formalism by Tanaka and Kashiwaya.\cite{Tanaka} The BTK kernel for
$d$-wave HTSC's in the case of pure $c$-axis tunneling without any
planar momentum components is given by
\begin{equation}
1 + |A|^2 - |B|^2 =
\frac{16(1+|\Gamma|^2)+4Z^2(1-|\Gamma^2|^2)}{|4+Z^2(1-\Gamma^2)|^2},
\end{equation}
where $\Gamma = E/\Delta(T,\phi)-\sqrt{\left( E/\Delta(T,\phi)
\right)^2-1}$. To account for the $d$-wave OP of Bi2212
$\Delta(T,\phi)=\Delta(T) \cos(2\phi)$ was chosen with $\phi$ as
the azimuthal angle in the Cu-O layer.\cite{Won} For the
temperature dependence of OP magnitude of the underdoped crystal
used the following empirical approximation \cite{gapftn} was
adopted.
\begin{equation}
\Delta(T) = \Delta(0) \tanh(3.2\sqrt{T_c/T - 1}),
\end{equation}
The above expression for the gap of the underdoped Bi2212 crystal
was obtained by fitting the temperature dependence of the gap used
in Ref. 31 
to the data in Ref. 25{ } 
for underdoped Bi2212 crystals with $T_c$=83 K, where we used
\cite{Oda} $\Delta(0)\approx$ 40 meV as the maximum value of the
OP at zero temperature. Equation (2) agrees very well with the
numerical solution of the BCS gap function except near $T_c$,
where underdoped samples are known to render a larger
gap.\cite{Miya1,Miya2} Using these parameters the normalized
conductance was calculated according to the BTK formula \cite{BTK}
as
\begin{equation}
\sigma(E) = \frac{4}{4+Z^2} \int^{\infty}_{-\infty} \frac{1}{2\pi}
\int^{2\pi}_0 [1 + |A|^2 - |B|^2] d\phi \times [-\frac{\partial
f(E-eV)}{\partial eV}]dE,
\end{equation}
where $f(E)$ is the Fermi distribution function. The values of the
barrier strength parameter $Z$ at different temperatures were
chosen by fixing the calculated conductance of Eq. (3) for $V$=0
to the measured differential conductance in Fig. 2. As shown in
the inset of Fig. 3 the values of $Z$ determined in this way are
proportional to the logarithm of $R_c(T)$, which is consistent
with earlier observations.\cite{Chuang} In addition, the
temperature dependence of $Z$ agrees with the scenario of
gradually weakening barrier strength with increasing temperature.
Figure 3 displays the results of Eq. (3) corresponding to the
$dI/dV$ curves of Fig. (2) for $T$=52.1$-$79.6 K. The overall
qualitative features of the calculated normalized conductance
curves have almost one to one correspondence to the experimental
data. At low temperatures, where the $Z$ value is supposed to be
relatively large, the calculated curves show gap-like features
which are qualitatively similar to the observed ones. With
increasing temperature the barrier strength parameter $Z$
decreases and both the calculated and experimental conductance
curves develop an AR-like enhancement near the zero-bias voltage
in a continuous manner. The most pronounced discrepancy between
the two sets of curves, however, is the width of the AR
enhancement at higher temperatures; the measured widths are
significantly narrower than the calculated ones. The discrepancy
was caused mainly by the above-mentioned premature switching of a
junction state to the quasiparticle branch; the switching took
place before the voltage reached the gap edge and thus the width
of the conductance enhancement was made narrower. Other
possibilities are the decrease of the AR process itself due to
thermal excitation of sub-gap-energy particles to the state above
the gap and the thermal smearing of the gap edge at temperatures
very close to $T_c$. Both effects may decrease the effective OP
magnitude experienced by the quasiparticles to a value smaller
than that of Eq. (2).

We also observed similar gradual reduction of the barrier
potential $Z$ in a mesa (OD1) fabricated on an as-grown overdoped
crystal (data are not shown). At temperatures sufficiently below
$T_c$ ($\approx$ 89 K) the IVC exhibited typical quasiparticle
branches with high hysteresis. Thus, in the low temperature range,
intrinsic junctions behaved as tunneling junctions with a high
barrier potential. Approaching $T_c$, however, the hysteresis
gradually disappeared and the IVC became more or less SNS-junction
like. This strongly suggests that the intrinsic junctions near
$T_c$ may satisfy the clean interface condition and one may be
able to observe the ZBCE by the AR in the c-axis conduction
measurements.

Taking the presence of $c$-axis AR for granted near $T_c$, we
carefully investigated the existence of the $c$-axis ZBCE above
$T_c$ of mesas fabricated on single crystals of varied doping
levels. The inset of Fig. 4(b) shows the resistive transition of
another overdoped mesa OD3, taken in a four-terminal
configuration. The mesa contained only several IJJ's. The
superconducting transition is relatively sharp and the temperature
of maximum $R_c(T)$, which should be close to $T_c$,\cite{Tc} is
about 90.3 K. Since measurements in a four-terminal configuration
do not include the surface effect, $R_c(T)$ does not exhibit a
finite resistance below $T_c$ as in Fig. 1. Thus, the inset of
Fig. 4(b) illustrates the intrinsic transition properties of the
inner junctions. As will be described below, the two arrows denote
the temperatures above which the ZBCE in the four- (left arrow)
and three-terminal (right arrow) configurations disappears.

Fig. 4(a) illustrates the differential conductance measured in the
four-terminal configuration at temperatures very close to $T_c$.
The zero-bias peak at each temperature was caused by the Josephson
pair tunneling over the stacked intrinsic junctions. With slightly
increasing temperature the conductance peak reduces rapidly and
disappears completely at 90.9 K, which is defined as the intrinsic
value of $T_c$ of the crystal. We notice that the value of $T_c$
turns out to be slightly higher than the temperature of maximum
$R_c(T)$. The concave background conductance represents the
existence of the pseudogap in this temperature range.\cite
{Parker} The conductance data measured in a three-terminal
configuration from the same central mesa are shown in Fig. 4(b).
One should note that in this case the properties of the junctions
in one of the smaller mesas including the surface junction were
measured. At 90.9 K, where the pair-tunneling zero-bias
conductance peak completely vanishes in the corresponding
four-terminal measurements, a clear ZBCE still persists, although
its magnitude is much smaller than the four-terminal counterpart.
With increasing temperature the ZBCE gradually reduces and
disappears completely only around 91.9 K. Thus, in this
three-terminal configuration, one has a 1-K temperature window of
ZBCE above the $T_c$ measured by the four-terminal configuration.
In a different mesa (OD2) fabricated on another overdoped crystal,
a 2-K temperature window of ZBCE above $T_c$ was obtained (data
not shown). We believe that the ZBCE was caused by an AR in the
surface junction between the normal electrode consisting of the Au
film and the surface Cu-O bilayer with suppressed
superconductivity, and the "superconducting" electrode of the
first inner Cu-O bilayer. Since the inner Cu-O bilayer is no more
superconducting above 90.9 K, the AR effect in Fig. 4(b) is either
from the thermally fluctuating superconducting order \cite {fluct}
or from the proposed preformed pairs in the pseudogap state. Then
the significantly reduced magnitude of the ZBCE in the
three-terminal configuration in Fig. 4(b) compared to the one in
the four-terminal configuration in Fig. 4(a) is easily explicable,
because both the thermally fluctuating pairs and the preformed
pairs in the pseudogap region are in phase-incoherent states.

If the AR was genuinely from the preformed pairs as suggested by
Ref. 13, the temperature window of the three-terminal ZBCE above
$T_c$ should get wider for underdoped crystals. In the
three-terminal configuration in Fig. 2, the ZBCE in the mesa of
UD2 disappeared at 82.1 K, which was in fact even lower than the
transition temperature ($T_c$=83.5 K) determined from the
four-terminal conductance measurements (not shown) on the same
crystal. We suppose that this unphysical behavior obtained in UD2
was caused by the possible doping inhomogeneity along the depth of
the mesa (lower doping near the surface of crystals), which was
introduced during the annealing process of lowering the doping
level. Due to this doping inhomogeneity the larger lower mesa
probed by the four-terminal configuration may have shown even
higher temperature of disappearing ZBCE than the upper mesa probed
by three-terminal configuration. No doping inhomogeneity is
supposed to develop in as-grown overdoped crystals. Since in UD2
the top Cu-O bilayer of the lower mesa was located only six layers
below the surface Cu-O bilayer,\cite{inhomogeneity} however, the
effect of doping inhomogeneity should have been marginal. Thus the
lack of the conductance enhancement in this underdoped crystal in
the three-terminal configuration cannot be explained by the doping
inhomogeneity only. We conclude that, contrary to the theoretical
expectation, we have observed no appreciable ZBCE above $T_c$ in
the underdoped crystal in the three-terminal configuration.

The great advantage of this measurement technique is in the fact
the three- and four-terminal configuraions were prepared on the
same "central" mesa in an identical piece of single crystal. This
geometry enabled us to determine the temperature, $T_c$, where the
pair tunneling ceases to exist in a four-terminal configuration.
Then a delicate change of the temperature near $T_c$ in the
three-terminal configuration, where the ZBCE due to the possible
AR from phase-incoherent Cooper pairs starts disappearing, can be
monitored very accurately.

In summary, we report that an AR has been observed in Au$/$Bi2212
single crystal junctions with the transport current along the $c$
axis of the Bi-2212 HTSC's, in both underdoped and overdoped
states. The features of the differential conductance curves
changed from gap-like depression at low temperatures to an AR-like
enhancement near $T_c$. We attribute such continuous evolution to
the AR between the proximity-suppressed surface Cu-O bilayer and
the superconducting inner one as the barrier potential decreases
with the temperature approaching $T_c$ from below. We utilized
this appearance of the AR effect near $T_c$ to investigate the
existence of the preformed pairs in the psudogap state above
$T_c$. In as-grown overdoped mesas we have observed maximally a
2-K temperature range where the three-terminal measurements
exhibit the AR-induced ZBCE above $T_c$ which was determined in a
four-terminal configuration. On the other hand, no appreciable
ZBCE above $T_c$ has been detected in single crystals which were
deep in the underdoped state, although theoretical consideration
suggests a wider temperature range of ZBCE above $T_c$. In the
light of the results from underdoped crystals the ZBCE itself
above $T_c$ in overdoped crystals may have been the thermal
fluctuation effect of the superconducting order. If that is the
case we have observed in this work no ZBCE by the AR from
preformed pairs in the pseudogap state. Our results then agree
with the recent report \cite{Dagan} which negates the AR effect in
$ab$-planar junctions by the preformed pairs in the pseudogap
state but are in disaccordance with the observation of
thermoelectrically induced vortices \cite{Nernst} in a wide
temperature range above $T_c$ in underdoped
La$_{2-x}$Sr$_x$CuO$_4$. We believe that, more rigorous
examination is also required on the validity of the suggested
observation\cite{Choi} of the AR effect from phase-incoherent
preformed pairs.

\section*{Acknowledgements}
One of us (H.-J. Lee) appreciates helpful private communications
with Y. Bang and H.-Y. Choi, and Y. Bang's critical reading of the
manuscript. Useful discussion with Y.-J. Doh is also appreciated.
The underdoped single crystals were provided by N. Momono and M.
Ido in Hokkaido University. This work was supported by Korea
Research Foundation, the center for excellency (SRC) administered
by Korea Science and Engineering Foundation, and Basic Science
Research Institute in POSTECH.


\begin{figure}[p]
\begin{center}\leavevmode
\includegraphics[width=0.6\linewidth]{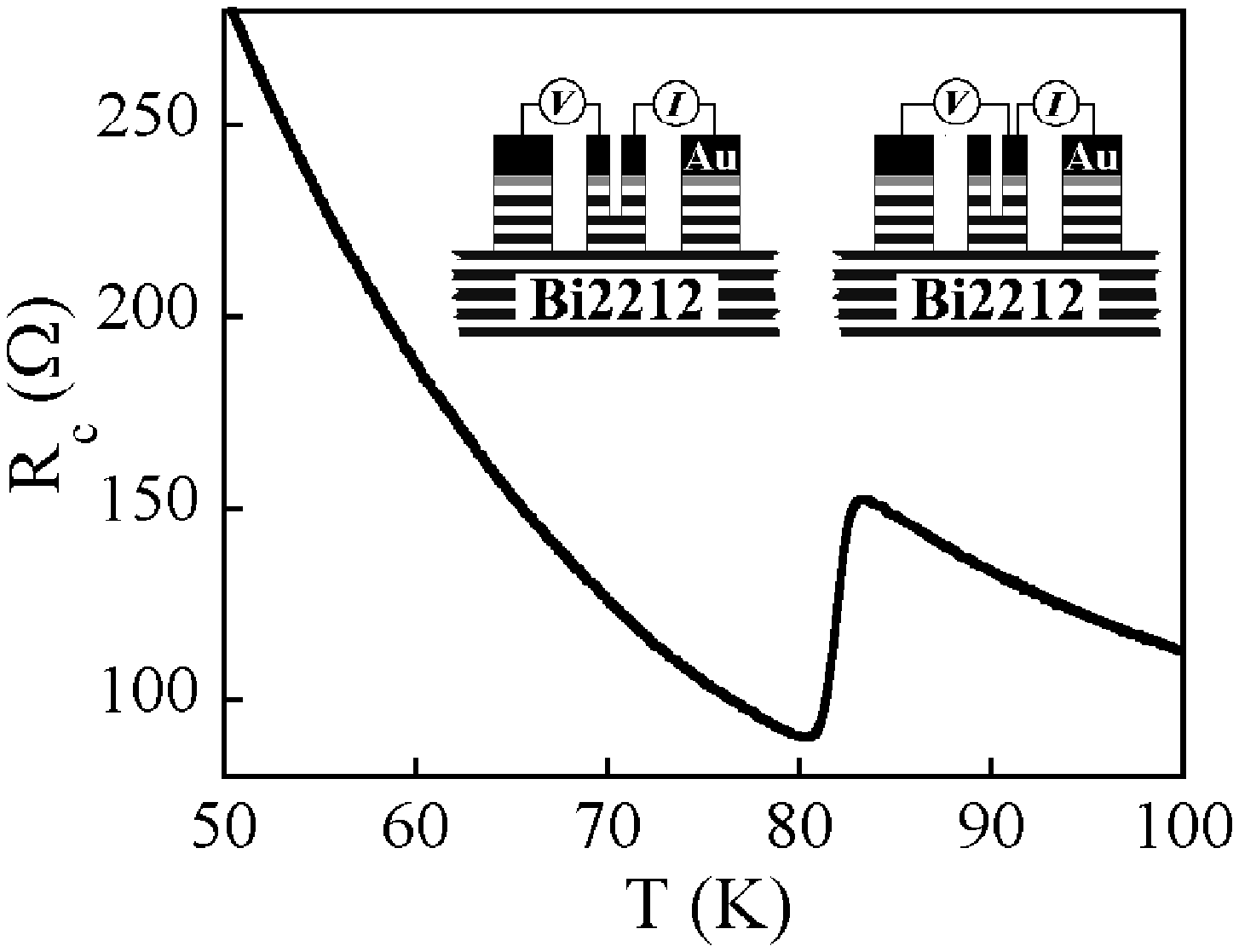}
\caption{Typical $R_c(T)$ curve measured in a 3-terminal
configuration of the mesa UD2, which was prepared on an underdoped
Bi2212 single crystal. The maximum of $R_c(T)$ takes place at
about 82.5 K ($\approx T_c$). The finite resistance below $T_c$
was from the surface junction, where no superconducting transition
was detected down to $\sim$4.2 K. Inset: schematic view of the
sample geometry and the configuration for four-terminal (left one)
and three-terminal (right one) measurements.}
\label{fig1}\end{center}\end{figure}

\begin{figure}[p]
\begin{center}\leavevmode
\includegraphics[width=0.6\linewidth]{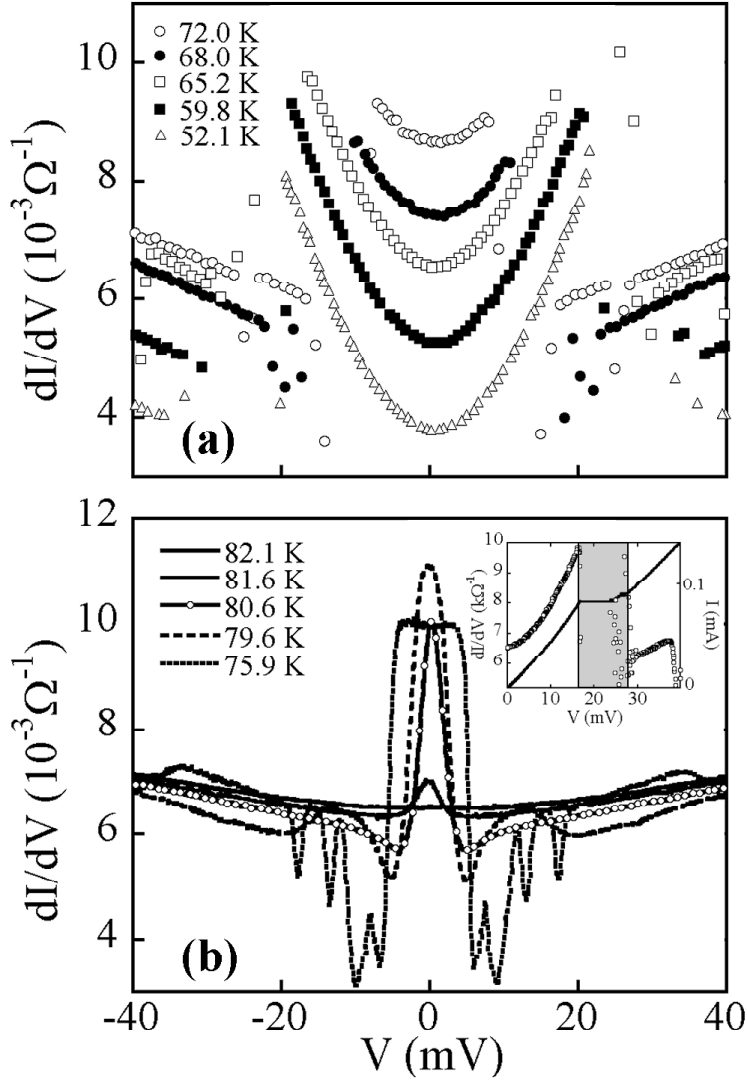}
\caption{Differential conductance of the NID surface junction (a)
for a relatively low temperature range and (b) for a high
temperature range near $T_c$. The gap-like depression of the
conductance at low temperatures gradually develops into AR-like
enhancement in a continuous manner with increasing temperature. At
$T\approx$ 82.1 K and above no conductance enhancement around zero
bias was detected. Inset: IVC ($-$), and the differential
conductance ($\circ$) {\it vs} voltage at $T\approx$ 65.2 K. The
shaded area marks the region of the voltage jump due to the
premature switching of the surface junction to the quasiparticle
branch as discussed in the text.}
\label{fig2}\end{center}\end{figure}

\begin{figure}[p]
\begin{center}\leavevmode
\includegraphics[width=0.6\linewidth]{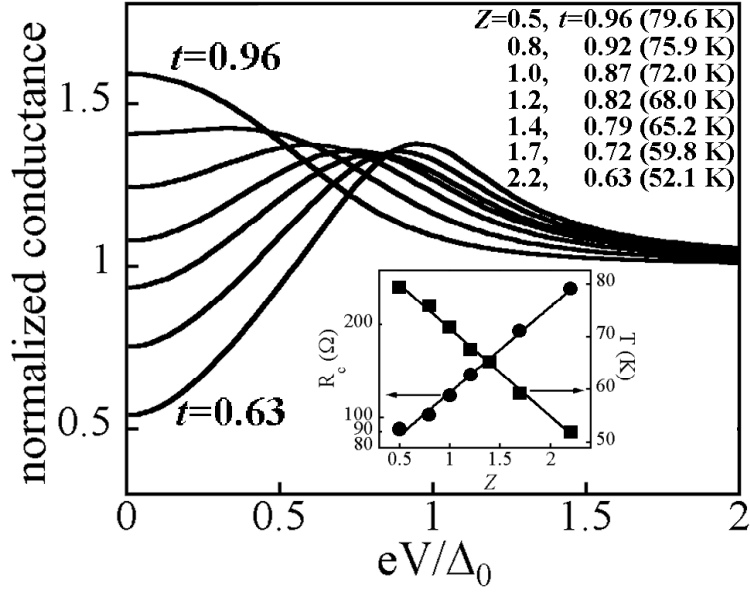}
\caption{Calculated normalized conductance using Eq. 3 based on
the $d$-wave BTK formalism. Fixing the zero-bias values of
conductance at various temperatures to the corresponding observed
ones in Fig. 2, continuously decreasing parameter value of the
barrier strength $Z$ gives the best qualitative fit to the results
in Fig. 2. Inset: the temperature dependence of $Z$
($\blacksquare$) and the $Z$ dependence of the $c$-axis resistance
$R_c(T)$ ($\bullet$). $Z$ is almost linearly proportional to
$\log(R_c(T))$ and $T$. Lines are guides to the eye.}
\label{fig3}\end{center}\end{figure}

\begin{figure}[p]
\begin{center}\leavevmode
\includegraphics[width=0.6\linewidth]{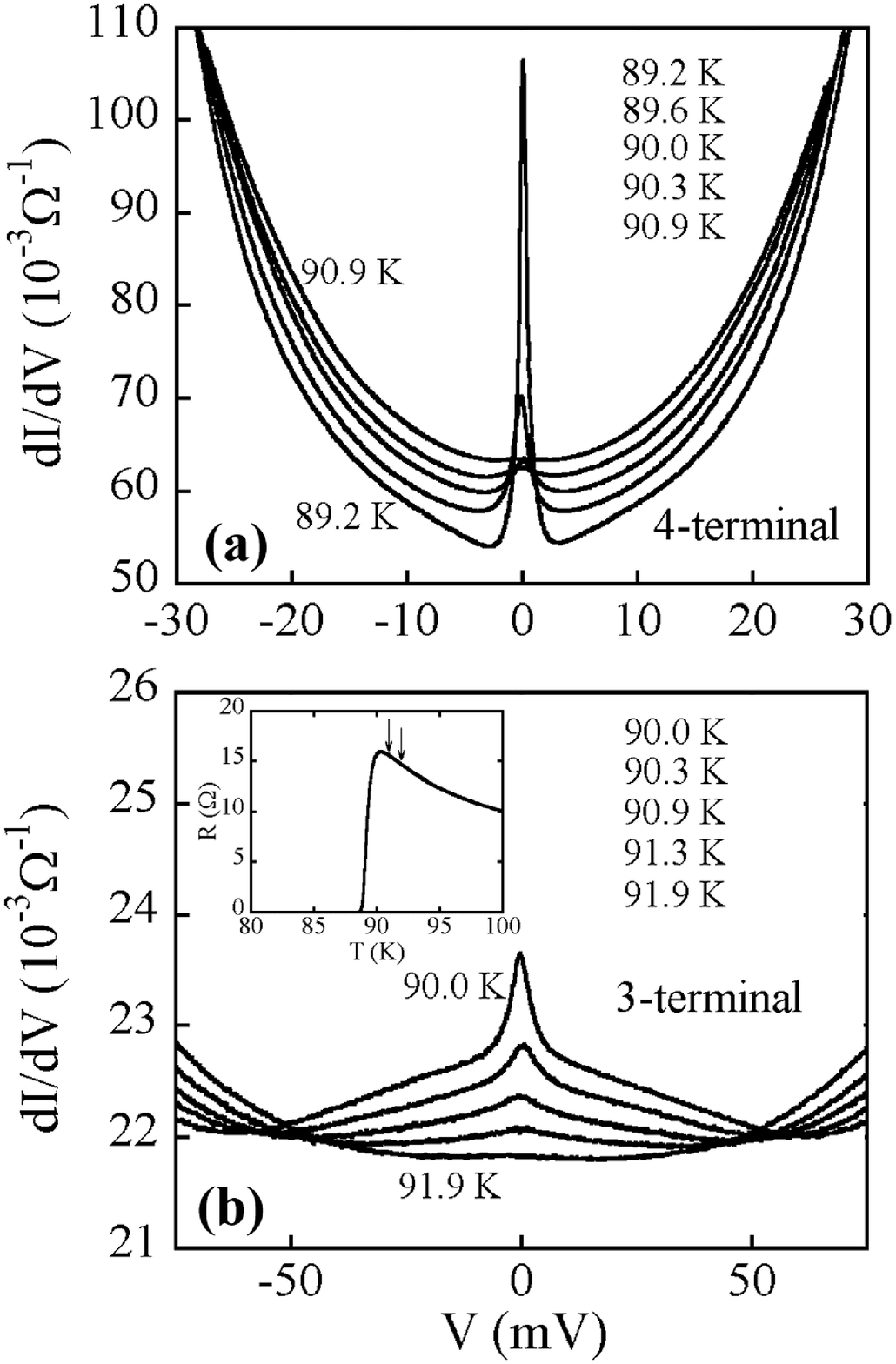}
\caption{(a) The differential conductance of the mesa OD3 measured
in the four-terminal configuration at temperatures very close to
$T_c$. (b) The differential conductance of the same sample as used
in (a) but in  a three-terminal configuration. Inset: the
resistive transition of the overdoped mesa OD3 obtained in a
four-terminal configuration, which represents the intrinsic
properties of the inner junctions. The two arrows denote the
temperatures above which the ZBCE obtained in the four- (the left
one) and three-terminal (the right one) configurations
disappears.} \label{fig4}\end{center}\end{figure}


\begin{thebibliography}{00}

\bibitem{DJ} D. J. Van Harlingen, Rev. Mod. Phys. {\bf 67}, 515 (1995).

\bibitem{twin} K. A. Kouznetsov, A. G. Sun, B. Chen, A. S. Katz, S. R. Bahcall, John Clarke,
R. C. Dynes, D. A. Gajewski, S. H. Han, M. B. Maple, J.
Giapintzakis, J.-T. Kim, and D. M. Ginsberg, Phys. Rev. Lett. {\bf
79}, 3050 (1997).

\bibitem{Ng} S. Sinha and K.-W. Ng, Phys. Rev. Lett. {\bf 80}, 1296 (1998).

\bibitem{Wei} J. Y. T. Wei, N.-C. Yeh, D. F. Garrigus, and M. Strasik, Phys. Rev. Lett.
{\bf 81}, 2542 (1998).

\bibitem{Covington} M. Covington, M. Aprili, E. Paraoanu, L. H. Greene,
F. Xu, J. Zhu, and C. A. Mirkin,  Phys. Rev. Lett. {\bf 79}, 277
(1997).

\bibitem{Mossle} M. M\"o\ss le and R. Kleiner, Phys. Rev. B {\bf
59}, 4486 (1999).

\bibitem{Schulz} R. R. Schulz, B. Chesca, B. Goetz, C. W. Schneider, A. Schmehl, H. Bielefeldt,
H. Hilgenkamp, J. Mannhart, and C. C. Tsuei, Appl. Phys. Lett.
{\bf 76}, 912 (2000); R. R. Schulz, B. Chesca, B. Goetz, C. W.
Schneider, A. Schmehl, H. Bielefeldt, H. Hilgenkamp, J. Mannhart,
and C. C. Tsuei, Physica C {\bf 341-348}, 1651 (2000).

\bibitem{Tanaka} Y. Tanaka and S. Kashiwaya, Phys. Rev. Lett. {\bf 74}, 3451 (1995).

\bibitem{Andreev} A. F. Andreev, Sov. Phys. JETP {\bf 19}, 1228
(1964).

\bibitem{BTK} G. E. Blonder, M. Tinkham, and T. M. Klapwijk, Phys. Rev. B {\bf
25}, 4515 (1982).

\bibitem{IJJ} R. Kleiner and P. M{\"u}ller, Phys. Rev. B {\bf 49},
1327 (1994); A. Yurgens, D. Winkler, N. V. Zavaritsky, and T.
Claeson, Phys. Rev. B {\bf 53}, R8887 (1996).

\bibitem{Kim} N. Kim, Y.-J. Doh, H.-S. Chang, and H.-J. Lee, Phys. Rev. B {\bf 59}, 14639 (1999).

\bibitem{Choi} H.-Y. Choi, Y. Bang, D. K. Campbell, Phys. Rev. B {\bf 61}, 9748
(2000).

\bibitem{Emery} V.J. Emery and S.A. Kivelson, Nature {\bf 374}, 434
(1995) and the references therein.

\bibitem{Dagan} Y. Dagan, A. Kohen, G. Deutscher, and A. Revcolevschi,
Phys. Rev. B {\bf 61}, 7012 (2000).

\bibitem{Kim2} N. Kim, Y.-J. Doh, H.-S. Chang, and H.-J. Lee, Physica C {\bf
341-348}, 1563 (2000).

\bibitem{Doh1} Y.-J. Doh, H.-J. Lee, and H.-S. Chang, Phys. Rev. B {\bf 61},
3620 (2000).

\bibitem{pTc} M. R. Presland, J. L. Tallon, R. G. Buckley, R. S. Liu,
and N. E. Flower, Physica C {\bf 176}, 95 (1991); G. V. M.
Williams, J. L. Tallon, R. Michalak, and R. Dupree, Phys. Rev. B
{\bf 54}, R6909 (1996).

\bibitem{Mossle2} M. M\"o\ss le, R. Kleiner, R. Gatt, M. Onellion, and P. M\"uller,
Physica C {\bf 341-348}, 1571 (2000).

\bibitem{Rowell} J. M. Rowell and W. L. McMillan, Phys. Rev. Lett.
{\bf 16}, 453 (1966).

\bibitem{Nesher} O. Nesher and G. Koren, Phys. Rev. B {\bf 60}, 9287 (1999).

\bibitem{cat1} S. Matsuo, M. Suzuki, X. G. Zheng and S. Tanaka, Physica C {\bf 282-287},
1497 (1997); H. J. Tao, F. Lu, and E. L. Wolf, {\it ibid.} {\bf
282-287} 1507 (1997).

\bibitem{cat2} Ch. Renner, B. Revaz, J.-Y. Genoud, K. Kadowaki, and {\O}. Fischer,
Phys. Rev. Lett. {\bf 80}, 149 (1998); Y. DeWilde, N. Miyakawa, P.
Guptasarma, M. Iavarone, L. Ozyuzer, J. F. Zasadzinski, P. Romano,
D. G. Hinks, C. Kendziora, G. W. Crabtree, and K. E. Gray, {\it
ibid.} {\bf 80}, 153 (1998); L. Ozyuzer, Z. Yusof, J. F.
Zasadzinski, R. Mogilevsky, D. G. Hinks, and K. E. Gray, Phys.
Rev. B {\bf 57}, R3245 (1998).

\bibitem{Miya1} N. Miyakawa,
J. F. Zasadzinski, L. Ozyuzer, P. Guptasarma, D. G. Hinks, C.
Kendziora, and K. E. Gray, Phys. Rev. Lett. {\bf 83}, 1018 (1999).

\bibitem{Miya2} N. Miyakawa J. F. Zasadzinski, L. Ozyuzer, P. Guptasarma, C.
Kendziora, T. Kaneko, D. G. Hinks, and K. E. Gray, Physica C {\bf
341-348}, 835 (2000).

\bibitem{Ishibashi} T. Ishibashi, K. Sato, and K. Sato, Physica C
{\bf 341-348}, 1615 (2000).

\bibitem{Sri} H. Srikanth and A. K. Raychaudhuri, Phys. Rev. B {\bf
45}, 383 (1992).

\bibitem{Chuang} C. H. Chuang and T. T. Chen, Physica C
{\bf 265}, 89 (1996).

\bibitem{SAR} T. Matsumoto, S. Choopun, and T. Kawai, Phys. Rev. B
{\bf 52}, 591 (1995); see also N. Tsuda, T. Arao, T. Hosokawa, Y.
Shiina, N. Matsuda, and D. Shimada, Physica C {\bf 282-287}, 1489
(1997).

\bibitem{Won} H. Won and K. Maki, Phys. Rev. B {\bf 49}, 1397 (1994).

\bibitem{gapftn} J. H. Xu, J. L. Shen, J. H. Miller, Jr., and C. S. Ting, Phys.
Rev. Lett. {\bf 73}, 2492 (1994).

\bibitem{Oda} M. Oda, T. Matsuzaki, and M. Ido, Physica C {\bf
341-348}, 847 (2000).

\bibitem{Tc} The accurate value of $T_c$ should be obtained by the
disappearing temperature of the differential conductance in the
four-terminal measurements. In the resistive transition, however,
the $T_c$ can be at best loosely defined.

\bibitem{Parker} I. F. G. Parker, C. E. Gough, M. Endres, P. J.
Thomas, G. Yang, and A. Yurgens, Proc. SPIE, {\bf3480}, 11 (1998).

\bibitem{fluct} The AR effect from the thermally fluctuating superconducting
order should be weaker than from the preformed pairs, if there is
any, since the preformed-pair state is represented by phase
fluctuations only while the thermally fluctuating state is subject
to fluctuations both in the amplitude and the phase of the
superconducting order (Refer to Q. Chen, I. Kosztin, B. Janko, and
K. Levin, Phys. Rev. Lett. {\bf 81}, 4708 (1998); Y. Yanase and K.
Yamada, J. Phys. Soc. Jpn. {\bf70}, 3664 (2000)).

\bibitem{inhomogeneity} The number of Cu-O bilayers in one of the upper
mesas was identified by counting the number of quasiparticle
branches in its hyteretic IVC at temperatures sufficiently below
$T_c$.

\bibitem{Nernst} Z. A. Xu, N. P. Ong, Y. Wang, T. Kakeshita, S. Uchida,
Nature {\bf 406}, 486 (2000).

\end{thebibliography}
\end{document}